\documentclass[reprint, superscriptaddress, amsmath,amssymb, aps, pra, floatfix]{revtex4-1}
\usepackage[T1]{fontenc} 
\usepackage{mathptmx}
\usepackage{graphicx}
\usepackage{dcolumn}
\usepackage{bm}
\usepackage[colorlinks, breaklinks, citecolor=blue,linkcolor=black,urlcolor=black]{hyperref}

\begin{document}

\title{Doppler cooling of buffer-gas-cooled Barium monofluoride molecules}

\author{Yuhe Zhang}
\author{Zixuan Zeng}
\author{Qian Liang}
\affiliation{%
Interdisciplinary Center of Quantum Information, State Key Laboratory of Modern Optical Instrumentation, and Zhejiang Province Key Laboratory of Quantum Technology and Device of Physics Department, Zhejiang University, Hangzhou 310027, China
}%
\author{Wenhao Bu}
\affiliation{%
Beijing Automation Control Equipment Institute, Beijing 100074, China
}%
\author{Bo Yan}
\email{yanbohang@zju.edu.cn}
\affiliation{%
Interdisciplinary Center of Quantum Information, State Key Laboratory of Modern Optical Instrumentation, and Zhejiang Province Key Laboratory of Quantum Technology and Device of Physics Department, Zhejiang University, Hangzhou 310027, China
}%

\date{\today}

\begin{abstract}

We demonstrate one-dimensional Doppler cooling of a beam of buffer-gas cooled Barium monofluoride (BaF) molecules. The dependences of the cooling efficiency with the laser detuning, the bias filed and the laser intensity are carefully measured. We numerical simulate our experiment with a Monte Carlo method, and find the theoretic predictions consists with our experimental data. This result represents a key step towards further cooling and trapping of BaF molecules.

\end{abstract}

\maketitle

\section{introduction}

Cold molecules have important applications in controlled chemistry , quantum simulations and precision measurement \cite{Carr2009, DeMille2002, Yan2013, Gadway2016,Micheli2006}. Although with much more complex energy levels than atomic case \cite{DiRosa2004, Stuhl2008}, polar molecules have been successfully laser cooled and trapped \cite{Barry2014,Collopy2018,Anderegg2019}. Previous experiments have demonstrated laser cooling of SrF \cite{Shuman2010}, CaF \cite{Zhelyazkova2014}, YO \cite{Hummon2013}, YbF \cite{Lim2018} , BaH \cite{McNally2020} and so on. And in last  ten years,  laser cooling technique has been extended to polyatomic molecules, such as, SrOH \cite{Kozyryev2017b}, CaOH \cite{Baum2020}, YbOH \cite{Augenbraun2020} and CaOCH$_3$ \cite{Mitra2020}. Besides, some other species, such as MgF \cite{Xu2016}, TlF \cite{Norrgard2017} and AlF \cite{Truppe2019} are under intensive exploration. 

Barium monofluoride (BaF) molecule has been considered as an attractive candidate for laser cooling for various reasons \cite{Chen2016,Xu2017,Kogel2021}. It is a heavy molecule, has a very high sensitivity for measuring the electric dipole moment (EDM) \cite{Cheng2016,Aggarwal2018}, and parity violation \cite{DeMille2008,Cahn2014,Altuntas2018,Haase2021}. Once it can be cooled and trapped, the accuracy of those measurements will be improved significantly. Thus, several groups are pushing on the direction of laser cooling of BaF. The high-resolution spectrum involved in laser cooling has been measured in a buffer-gas cell \cite{Bu2017, Albrecht2020}; The deflection of BaF molecular beam by lasers has been observed \cite{Chen2017}. But similar with BaH \cite{McNally2020} and YbF molecules \cite{Lim2018}, the laser cooling of BaF is more difficult than light molecules. Since it is heavier, the deceleration is smaller, more photons are needed to be scattered. At the same time, the vibrational branching ratios are less favorited \cite{Chen2016}, making the scattering of many photons to be harder. There is also a substable state ($\Delta$ state) with energy lower than the excited state of the cooling transition, which might also induce leakage to dark states \cite{BARROW1988, Albrecht2020}. With these, the demonstration of laser cooling of BaF is still lacking. 

Here we report an experimental observation of laser cooling of BaF molecule. In Sec. \ref{sec2}, the energy levels involved in laser cooling  are introduced. In Sec. \ref{sec3}, we describe the experiment setup. With the right detuning, both laser cooling and heating are observed. In Sec. \ref{sec4}, the cooling effect versus various experimental parameters, such as the detuning of lasers, the bias field and laser intensity are measured. We also numerically simulate our experiment with a Monte Carlo method, and find they agree with the experiment results quite well. Finally, we conclude our paper in Sec. \ref{sec5}.

\begin{figure}[]
\centering
\vspace{2mm}
\includegraphics[width=0.48\textwidth]{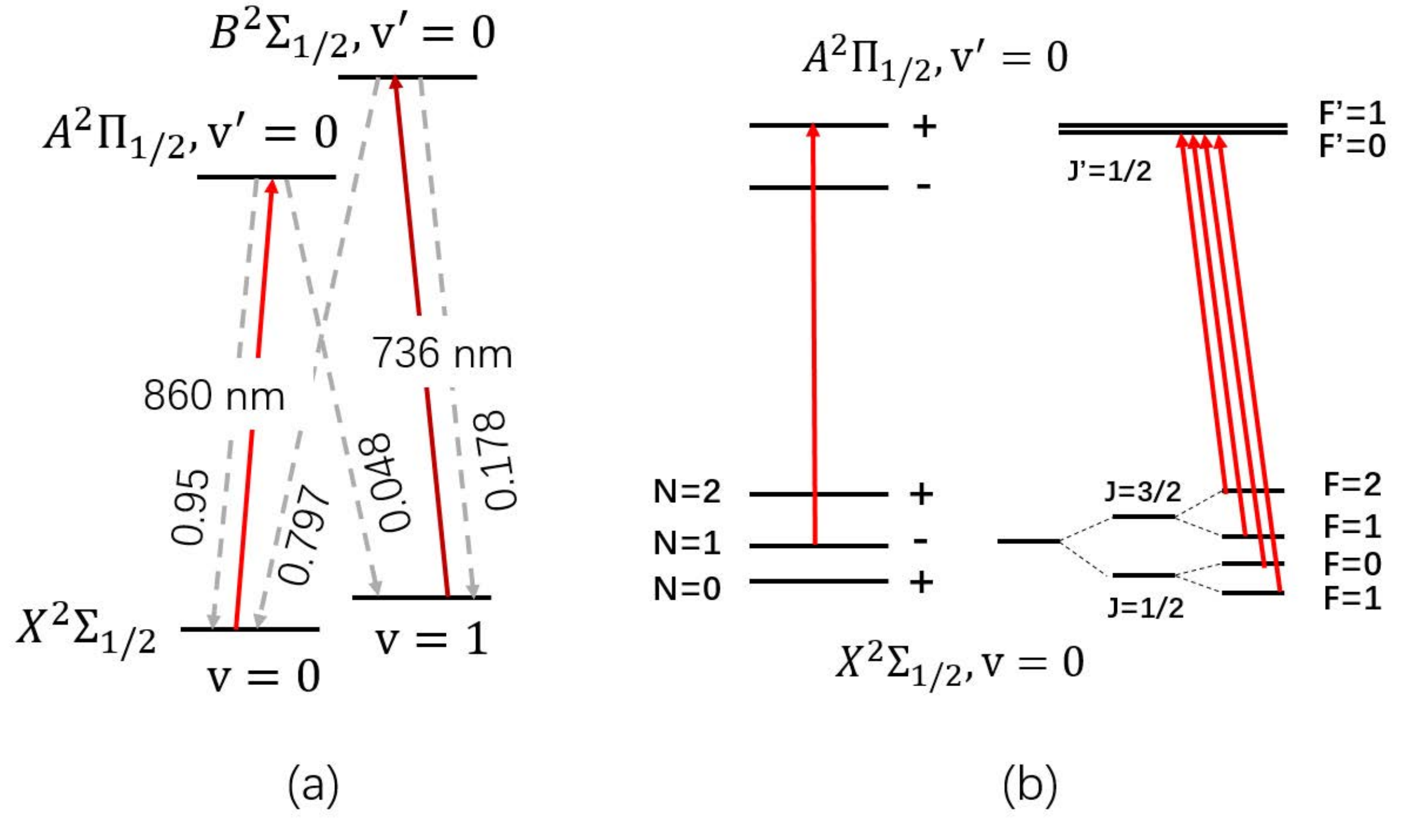}
\caption{(color online) \label{energy_level}
Energy levels involved in laser cooling of BaF molecules. (a) shows the electronic states and the vibrational states. The branching ratios are also included in the plot. A 736 nm laser is used to repump molecules in $v=1$ state back to $v=0$ state. But still, molecules will decay to dark states (such as $v=2$) with probability about $0.17\%$ . (b) shows the rotational states and hyperfine states involved in the cooling scheme. Multi-frequency lasers are used to cover the hyperfine splitting.
}
\end{figure}

\begin{figure*}[tbp]
\centering
\includegraphics[width=0.8\textwidth]{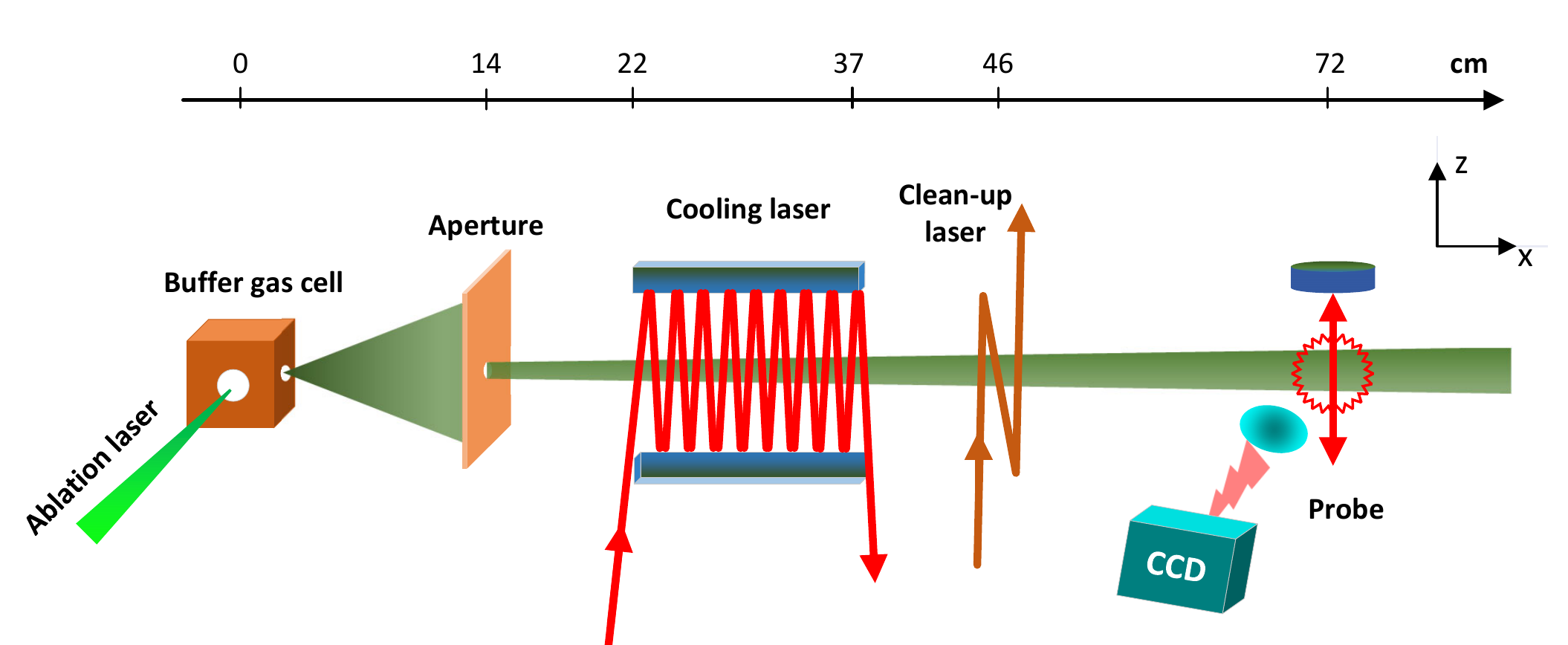}
\caption{(color online) \label{setup}
Experimental setup. The total working distance is about 70 cm. BaF molecules are created in a buffer gas cell and collimated by an aperture. Molecules are cooled in the cooling regime and finally imaged by a camera.
}
\end{figure*}
\section{Energy levels involved in laser cooling}\label{sec2} 
We are working on $^{138}$BaF molecule. The energy levels related with laser cooling are shown in Fig. \ref{energy_level}. The main cooling laser is near resonance with $|X^2\Sigma_{1/2}, ~v=0\rangle \to |A^2\Pi_{1/2}, ~v'=0\rangle, ~\lambda_{00}=859.8390$~nm. Molecules in the excited state $| A^2\Pi_{1/2}, ~v'=0\rangle$ will also decay to $v=1$ and $v=2$ of ground state with branching ratios of $4.8\%$ and $0.15\%$ respectively. In order to close the vibrational transitions, molecules in $|X^2\Sigma_{1/2}, ~v=1\rangle$ state is repumped with a laser near resonance with $|X^2\Sigma_{1/2}, v=1\rangle \to |B^2\Sigma_{1/2}, v'=0\rangle$, $\lambda_{10}=736.7047$ nm. Comparing with our previous repump scheme of $|X^2\Sigma_{1/2},~ v=1\rangle \to |A^2\Pi_{1/2},~ v'=0\rangle$~\cite{Chen2017}, this new repump scheme enables the cooling force to be increased by 50$\%$. With the repump laser, the leakage to other states (mainly $v=2$ of the ground state) is about $0.17\%$. So molecules can scatter few hundreds of photons before they decay to the dark states.

In order to close the rotational transition, we choose the $|N=1,-\rangle \leftrightarrow |J'=1/2, +\rangle$ transition for the ground state and excited state. In order to cover the hyperfine states, we use acoustic-optical modulators (AOM) to shift the frequency of cooling laser to produce multi-frequency components with relative frequencies of $\{0,~17,~123,~ 151\}$~MHz. For the repump laser, a 38 MHz modulated electro-optical modulator (EOM) is used to produce $\{\pm 1, \pm 2\}$ sidebands to cover the hyperfine splitting \cite{Chen2017}.

\section{experimental setup}\label{sec3}
The experimental setup is shown in Fig.~\ref{setup}. The BaF molecules are produced by ablating the BaF$_2$ target with a 532-nm pulsed ND: YAG laser (the repetition rate is $1~{\rm Hz}$) in a 4 K cell. BaF molecules are pre-cooled by the 4 K buffer gas and come out the cell with a forward velocity around 200 m/s through a 5 mm hole. The molecular beam is then collimated by another 5 mm diameter aperture. 

In the cooling regime, the molecular beam is cooled in the transverse direction (along $z$ as shown in Fig. 2). The cooling laser and repump laser are combined together before entering the cooling regime. In order to increase the interaction time, lasers are retro-reflected $2\times$14 times by two mirrors. The cooling laser has a total power of $P_0=$167 mW and the repump laser has a power of 310 mW before entering the viewport. Both beams have a full width half maximum (FWHM) $d=$2.4 mm. The viewport windows of the vacuum chamber are anti-reflection coated with less than $1\%$ reflection. But after continuous working for several days, the loss of laser power through the viewport windows increases due to the dust produced by the laser ablation. When we take data for this paper, about 50\% of the total power are lost after the whole trip. A small bias magnetic field $B$ is applied to remix the Zeeman dark states \cite{Berkeland2002}, the angle between the laser polarization and the bias field is labelled as $\theta$. 

\begin{figure}[tbp]
\centering
\includegraphics[width=0.48\textwidth]{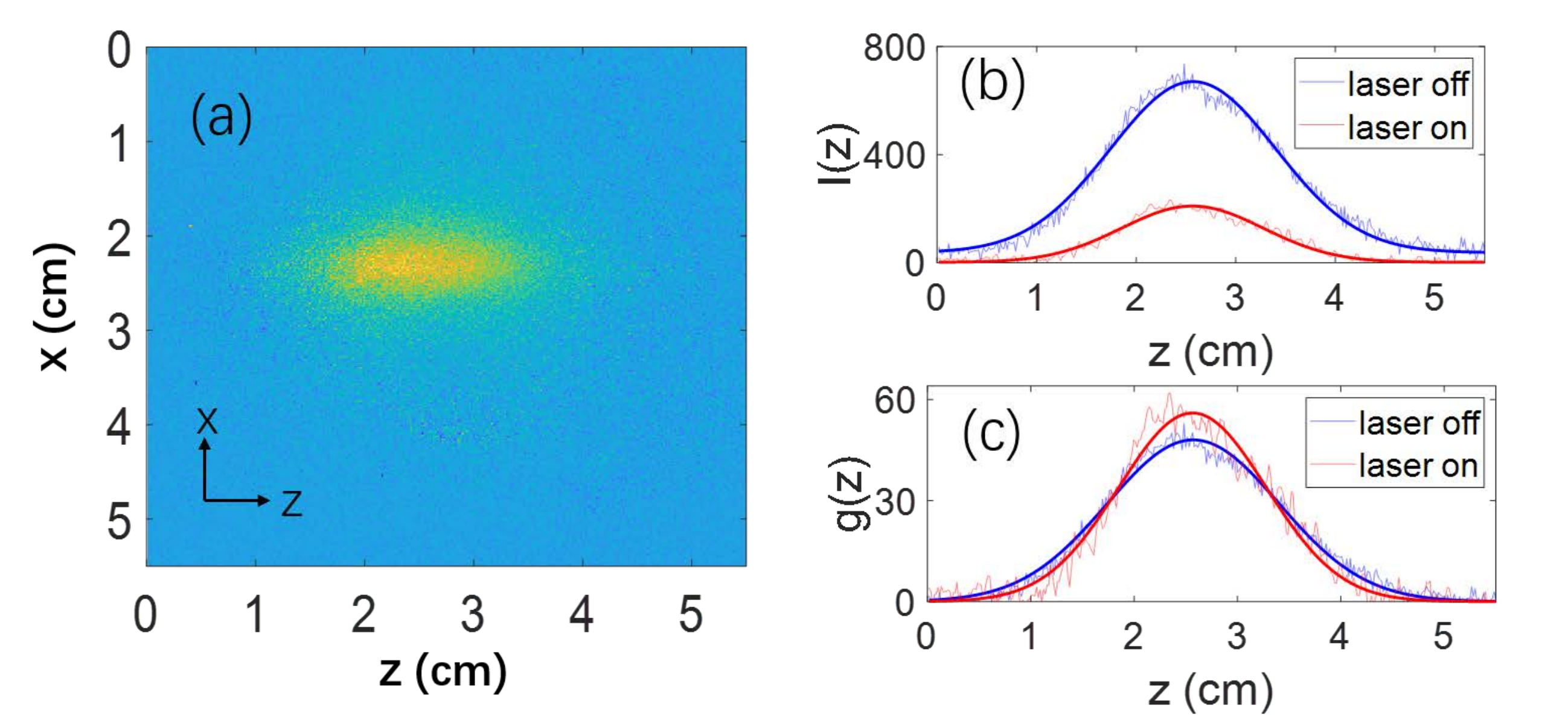}
\caption{(color online) \label{cooling}
(a) a typical image of the molecular beam. It has been averaged for 1500 times. (b) The molecule signal along $z$ direction. The smooth solid lines are fitting curves with function (\ref{Eq:fiting}). (c) shows the normalized results, the smooth solid lines are fitting curves with function (\ref{Eq:fittingnorm}). For all figure, the experimental parameters are $\delta=-5$ MHz, $B=0.82$ G, $\theta=70^\circ, P_0=167$~mW.
}
\end{figure}
After the cooling regime, molecules enter a clean-up regime. A repump laser (here we call it clean-up laser, with power 30 mW, beam size is the same as cooling beam) is used to pump molecules in $v=1$ of ground state back to $v=0$ of ground state before imaging. 
And finally, at the imaging regime, molecules are excited by a  probe beam, the florescence is recorded by a camera (CCD). From the imaging picture, we extract the information of the molecular beam size and flux. 

In a real experiment, we take two images at one experimental cycle, one is with molecular beam on (ablation laser on), one is without molecule. They are subtracted to get an image signal of molecular beam. Because the molecule flux actually is very low, we need to average several hundreds to get a reasonable good image. Figure \ref{cooling} (a) shows such a typical averaged fluorescence image of molecular beam. From the experimental setup of Fig. \ref{setup}, we can see the width of molecular image along $x$ direction is determined by the probe beam size. While the width along $z$ direction is determined by the transverse temperature of the molecular beam. In order to extract the information of the transverse temperature, we integrate the image along $x$ direction and fit it with a Gaussian function along $z$ direction as shown in Fig. \ref{cooling} (b), the fitting function is 
\begin{equation}\label{Eq:fiting}
I(z)=I_0+Ag(z) 
\end{equation}
where $g(z)$ is the normalized Gaussian function
\begin{equation}\label{Eq:fittingnorm}
g(z)=\frac{1}{\sqrt{\pi}R}e^{-(z-z_0)^2/R^2}
\end{equation}
The fitted size $R$ gives the information of transverse temperature of molecular beam. 

In order to reduce the long term drift, we actually take two kinds of molecular image at one experimental cycle. One is without cooling laser (fitted size is labelled as $R_0$), one is with cooling laser on (fitted size is labelled as $R$). The size ratio $R/R_0$ gives the information of cooling or heating. If $R/R_0<1$, it shows a cooling effect. If $R/R_0>1$, it shows a heating effect. Figure \ref{cooling} (b) and (c) shows such comparison when $P_0=167$ mW, $\delta=-5$~MHz, $B={ 0.82}$ G, $\theta=70^\circ$. Because there is still small probability for molecules decay to the dark states, the total molecular signal actually becomes smaller when the cooling laser is switched on (usually about 60\%-70\% molecule is depleted to dark states after passing through the cooling regime). But still we can extract the size ratio and clearly see there is a cooling effect from Fig. \ref{cooling}~(c). 
\begin{figure}[]
\centering
\includegraphics[width=0.4\textwidth]{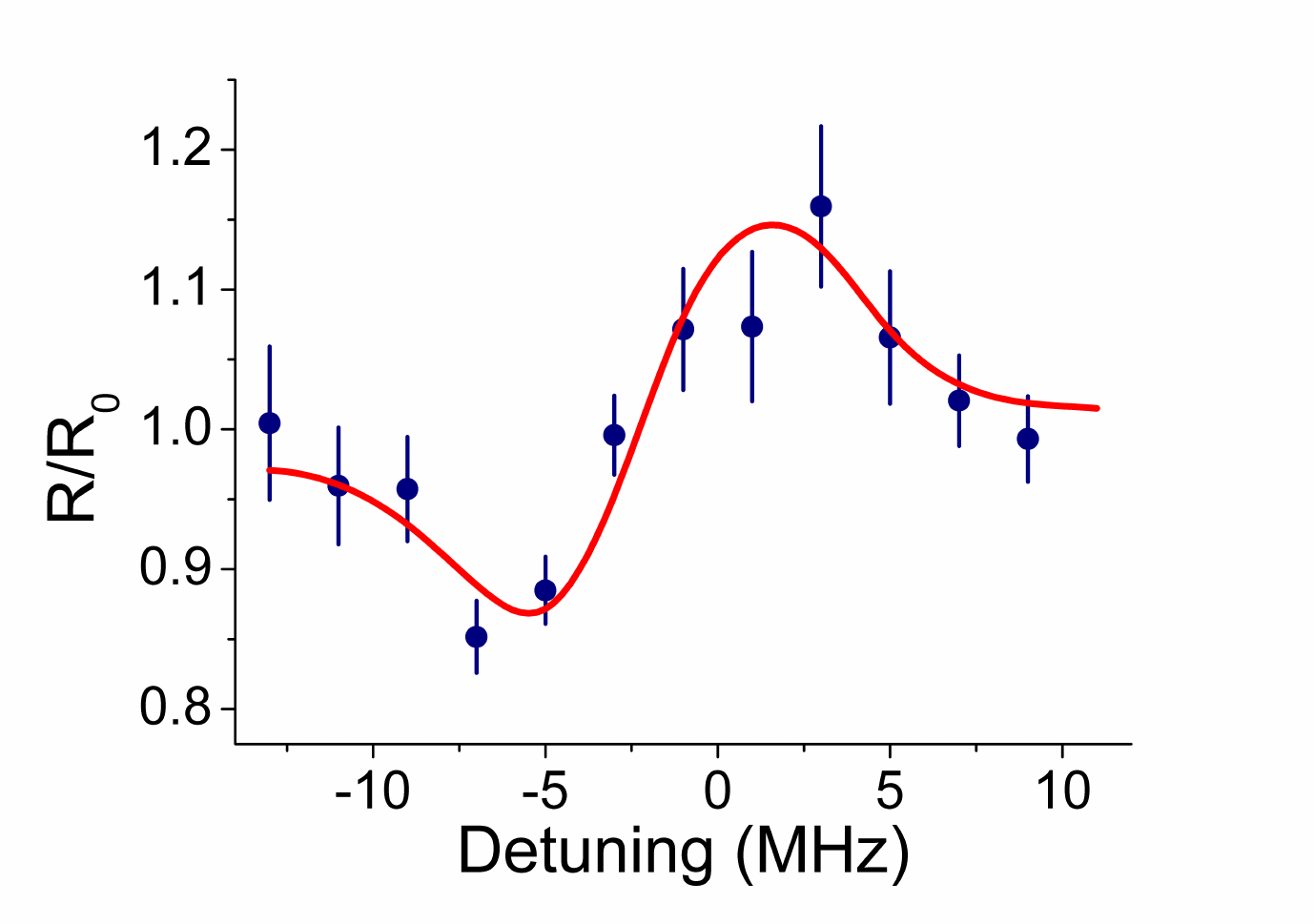}
\caption{(color online) \label{detuning_dep}
The dependence of cooling efficiency versus laser detuning. The dots are experimental data with $B=0.82$ G, $\theta=70^\circ, P_0=167$~mW, and the solid line come from numerical Monte Carlo simulation. Both cooling and heating show up depends on the detuning.  Error bars are standard deviations.
}
\end{figure}

\section{results and analysis}\label{sec4}
In order to carefully study the laser cooling effect for BaF molecule, we systemically change the detuning of the cooling laser, the bias magnetic field and the cooling laser power. And for comparison, we set up a simple Monte Carlo simulation of our experiment. 

The Doppler cooling force for BaF molecules is calculated with the multi-level master equation model \cite{Chen2017,Liang2019, Liang2021}. We only consider the main cooling laser and ignore the leakage channels to higher vibrational states and the $A'^2\Delta$ state, i.e., the ``4+12''-level model. Here we assume the four sidebands ideally address the four ground-state hyperfine sublevels with the same detuning $\delta$. The saturation parameters $s$ for all the sidebands are set  the same. 

\begin{figure}[]
\centering
\includegraphics[width=0.4\textwidth]{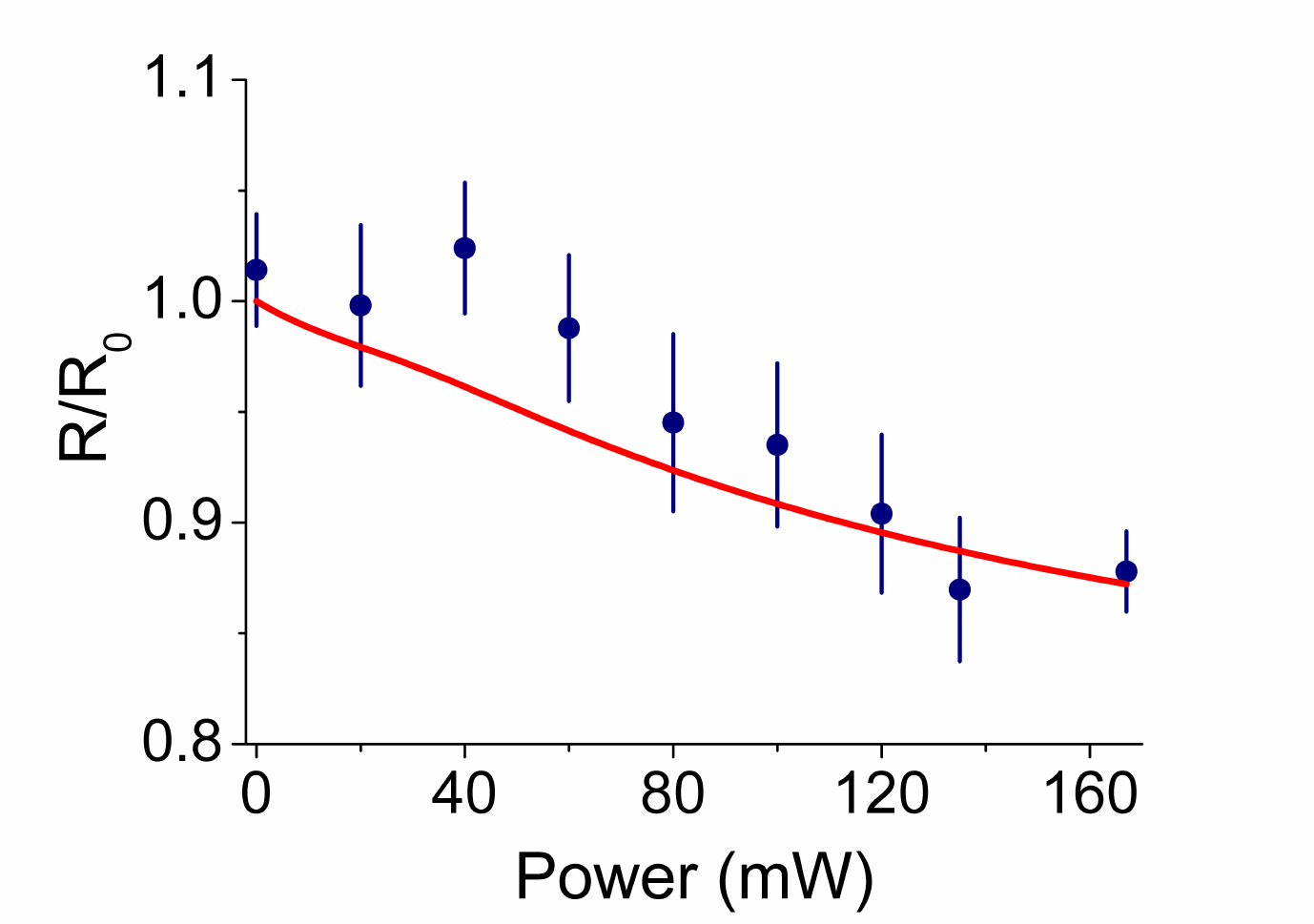}
\caption{(color online) \label{I_dep}
The dependence of the cooling efficiency on the incoming cooling laser power. The dots are experimental data with $\delta=-5$ MHz, $B=0.82$ G, $\theta=70^\circ$, and the solid line come from numerical Monte Carlo simulation. The cooling effect gets better when increasing the cooling power. Error bars are standard deviations.
}
\end{figure} 

We numerically simulate the trajectory of the molecules from the first source aperture to the detecting area. The initial conditions of these emitted molecules is sampled by the Monte Carlo method. According to the feature of buffer gas cooled molecular beam, the velocity distribution we used in the simulation is an equilibrated Maxwell distribution at a temperature of 4K plus a collective forward velocity of 200m/s. 
We use the experimental value, such as distances shown in Fig \ref{setup} to perform the Monte Carlo simulation. In order to make it simpler, we use only one averaged laser intensity and an effective cooling distance of 3 cm. We adjust the  laser intensity  to get the best fittings. From the fitting, we get   $s=12.5$ for $P_0=167$ mW. Consider the power and the size of laser beam, four-frequency components, the attenuation of the window. The fitted value of laser intensity is lower than the experimental value. This might come from the non-linear effect from a   tapered amplifier (TA). In order to get multi-frequency laser components, we use AOMs to shift the lasers first, and then combine them to inject  into a TA. Because the frequency differences between different components are about tens to one hundred MHz, it can cause higher-order frequencies and waist power to unwanted frequency components \cite{Ferrari1999,Barry2013}. 

Figure \ref{detuning_dep} shows the size ratio versus detuning when $P_0=167$ mW, $B=0.82$ G, and the angle between laser polarization and the remixed magnetic field $\theta=70^\circ$. It shows a clear detuning dependence of the cooling effect. When it is red detuned ($\delta <0$), a cooling effect is observed, the molecular beam size reduces 15\% around $\delta=-5$ MHz. On the other hand, when it is blue detuned ($\delta>0$), a heating effect shows up. The obvious cooling/heating window for detuning is about 3 MHz which is determined by the relative narrow linewidth of the excited state (about 2.8 MHz) and small hyperfine splitting (on the order of 10~MHz). The solid line is the numerical simulation of our experiment with $B=0.82$ G, $\theta=70^\circ$ and $s=12.5$. 

\begin{figure}[]
\centering
\includegraphics[width=0.43\textwidth]{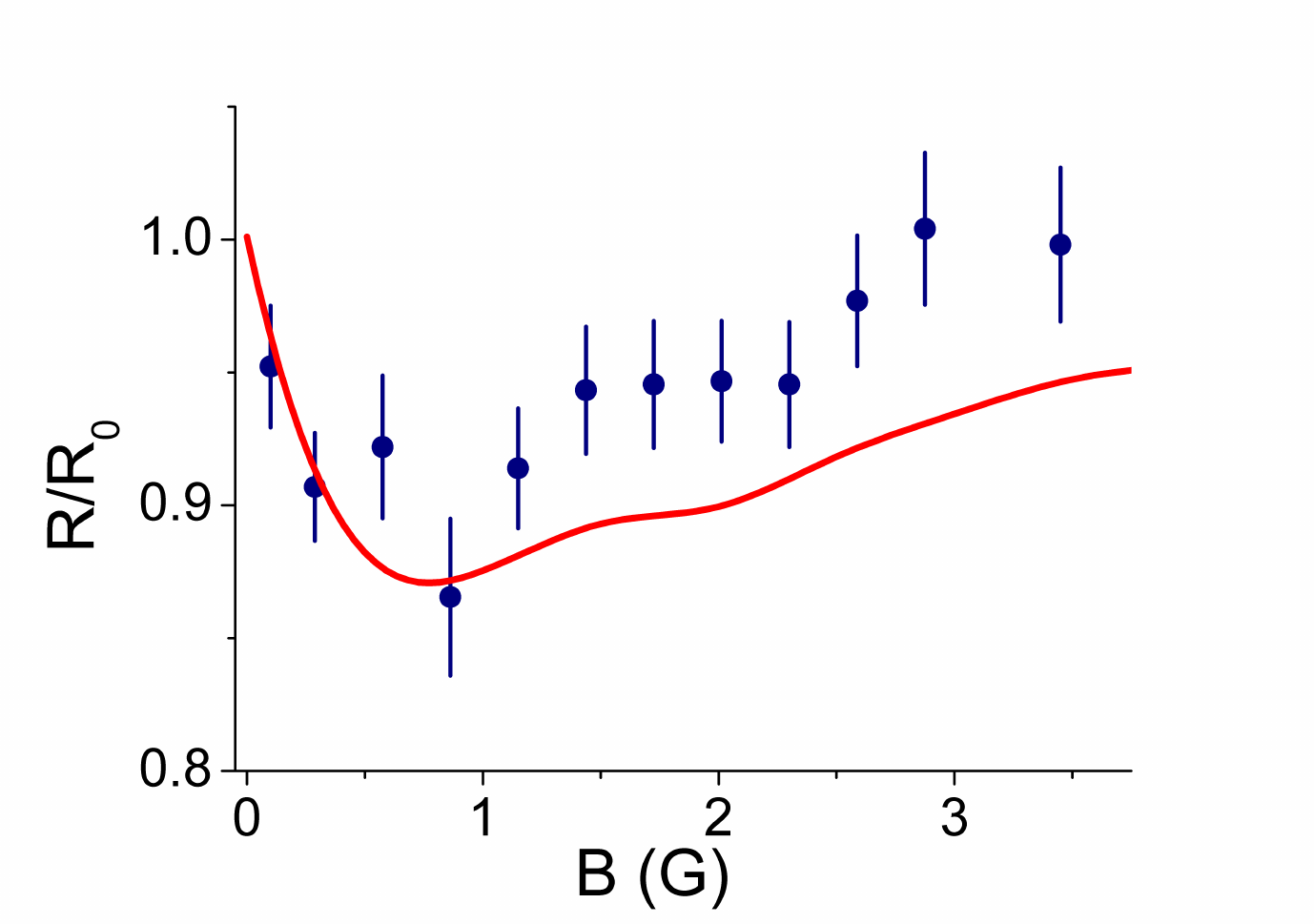}
\caption{(color online) \label{mag_dep}
The dependence of the cooling efficiency on the bias field. The dots are experimental data with  $P_0=167$ mW, $\delta=-5$ MHz and $\theta=70^\circ$, and the solid line come from numerical Monte Carlo  simulation. The best cooling appears around $B\simeq1$ G. Error bars are standard deviations.
}
\end{figure}

Figure \ref{I_dep} shows the size ratio versus the laser power. The experimental parameters are $\delta=-5$ MHz, $B=0.82$ G, $\theta=70^\circ$. As long as the laser power is not fully saturated, higher power gets better cooling effect. In our case, we cann't further increase the laser power, because the leakage to other dark states will be too high and the remaining molecular signal will be too lower to be detected.
The solid line comes from the numerical simulation with the experimental parameters, it captures the main feature of these data.

Figure \ref{mag_dep} shows the size ratio versus the bias field. The experimental parameters are $\delta=-5$ MHz, $P_0=167$ mW and $\theta=70^\circ$. The solid line represents the numerical simulation. When the remixed field is too low, the remixing of the Zeeman dark state doesn't work well, thus laser cooling process stops. On the other hand, if the magnetic field is too high, the Zeeman splitting mess up the cooling detuning and the cooling effect vanish. The best cooling regime happens around $B\simeq 1$~G. The cooling effect for experiment  is not as good as numerical simulation at higher bias field, it might due to the slight polarization rotation at each reflection from the mirror. The incoming cooling laser has a fixed polarization, but after 28 times reflection, the laser becomes slightly elliptic polarization. We find there are 30\% of the laser power at the orthogonal direction of the initial polarization direction after the laser are reflected 28 times. But overall, the numerical result captures the main trend of the experiment data.

\vspace{6mm}
\section{conclusion}\label{sec5}
To conclude, we have demonstrated the Doppler cooling of BaF molecules in one dimension. With the help of a new repump scheme, the cooling force increases and a clear cooling signal shows up. The dependence of cooling effect versus various experimental parameter are measured and compared with a numerical simulation. They consist quite well with each other.

On the other hand, the cooling effect is still limited by the number of photons can be scattered before molecules decay to dark states. The remaining leakage to dark states is about 0.17\%, it is still too high. With this leakage, it would be hard to realize sub-Doppler cooling, slowing and further magneto-optical trapping. In the future, we will try to further close the cycling transition by implement the repump of $v=2$ state, and also close the $\Delta$ state leakage \cite{Yeo2015,Collopy2018}.

\begin{acknowledgments}

We acknowledge the support from the National Key Research and Development Program of China under Grant No.2018YFA0307200, the National Natural Science Foundation of China under Grant Nos. U21A20437 and 12074337, Natural Science Foundation of Zhejiang province under Grant No. LR21A040002, Zhejiang Province Plan for Science and technology No. 2020C01019 and the Fundamental Research Funds for the Central Universities under No.2021FZZX001-02	.
\end{acknowledgments}

\bibliographystyle{apsrev4-1}
\bibliography{coolingbio}

\end{document}